\documentstyle[epsf,prl,preprint,aps]{revtex}

\tightenlines
\begin{document}
\draft 

\title{Large Magnetoresistance Ratio in Ferromagnetic
Single-Electron Transistors in the Strong Tunneling Regime}

\author{X. H. Wang}
\address{Department of Theoretical Physics, Lund University, \\
Helgonav{\"a}gen 5, S-223 62 Lund, Sweden}
\author{Arne Brataas$^{\dag}$}

\address{Department of Applied Physics and Delft Institute of
Microelectronics and Submicrontechnology (DIMES), \\
Delft University
of Technology, Lorentzweg 1, 2628 CJ Delft, The Netherlands}

\date{\today}
\maketitle

\begin{abstract}
We study transport through a ferromagnetic single-electron transistor. The
resistance is represented as a path integral, so that systems where the
tunnel resistances are smaller than the quantum resistance can be
investigated. Beyond the low order sequential tunneling and co-tunneling
regimes, a large magnetoresistance ratio at sufficiently low temperatures is
found. In the opposite limit, when the thermal energy is larger than the
charging energy, the magnetoresistance ratio is only slightly enhanced.
\end{abstract}


Electron tunneling between two ferromagnetic metallic leads separated by an
insulating layer has been studied since the seventies \cite{jul75}. The
novel feature of these structures is that the conductance depends on the
relative orientation of the magnetizations in the leads. By tuning the
external magnetic field, the magnetization of a ferromagnetic metal can be
controlled, and hence the conductance of the device changes. The influence
of the applied magnetic field on the transport properties is characterized
by the tunnel magnetoresistance ratio (TMR): 
\begin{equation}
\gamma =(R_{\text{AP}}-R_{\text{P}})/R_{\text{P}}\,,  \label{tmr}
\end{equation}
where $R_{\text{P}}$ ($R_{\text{AP}}$) is the resistance when the
magnetizations in the leads are parallel (anti-parallel). Devices with large
TMR can be used\ in future magneto-resistive sensors or magnetic random
access memories \cite{moo95}.

In the past years, reproducible measurements of TMR have been reported
\cite {moo95,Levy99,ono97,sche97,san97} in structures with ultrasmall
tunnel junctions.  When the capacitances of the tunnel junctions are
so small that the charging energy is larger than the thermal energy,
the current flow is blocked when the bias voltage is smaller than a
critical value determined by the junction capacitances. Such a Coulomb
blockade of electron tunneling through insulating barriers has been
widely investigated when the leads are nonmagnetic
\cite{ave90,ave91,schon,wang}. The single-electron transistor consists
of a metallic island connected to two metallic leads via tunnel
junctions and coupled to a gate-voltage $U_{g}$ via a capacitor
$C_{g}$. The lowest order perturbation in terms of the dimensionless
tunnel conductance $ \alpha =R_{K}/R_{T}$, where $R_{K}\equiv h/e^{2}$
is the quantum resistance and $R_{T}$ is the tunnel resistance,
predicts that when the thermal energy $k_{B}T$ is much smaller than
the charging energy $E_{c}=e^{2}/2C$, the conductance of the SET
vanishes for $C_{g}U_{g}/e=n$, and the conductance attains a maximum
for $C_{g}U_{g}/e=(n+1/2)$, where $n$ is an integral number
\cite{ave91}. When the tunnel resistances $R_{T}$ are not much smaller
than the quantum resistance, the Coulomb blockade of electron
tunneling is softened due to co-tunneling and higher order tunneling
processes \cite{ave90}.

In order to describe ferromagnetic single-electron transistors
(FSET's), the spin dependence of the density of states and the
transition matrix elements must be taken into account. Theoretical
works on this subject based on the lowest order sequential tunnelling
processes have recently been carried out \cite{bar98,bra99,kor99}. By
including the co-tunneling processes in the Coulomb blockade regime,
it has been shown \cite{tak98,ima99} that the TMR is larger in such
ultrasmall structures than in macroscopic devices, where the charging
energy is negligible. When the junction resistances are smaller than
the quantum resistance, the calculations based on sequential tunneling
or co-tunneling become inaccurate because higher order tunneling
processes determine the transport properties. It is so far unclear if
the TMR of the FSET is further enhanced by the higher order tunneling
processes.  This question cannot be addressed from the results of the
sequential tunneling or co-tunneling formulations. We will employ a
non-perturbative approach to clarify this issue and show that indeed
there is a further large enhancement of the TMR in the strong tunneling 
regime as compared with the weak tunneling case where the junction
resistances are larger than the quantum resistance. This enhancement
of the TMR is much larger than what calculations based on the
co-tunneling processes predict.

We consider the situation when the two leads are made of the same material
and have the same magnetization directions, and denote the resistance of the
FSET as $R_{\eta }$, where $\eta =\text{P}$ ($\eta =\text{AP}$) for the case
that the magnetization of the central grain is parallel (antiparallel) to
the magnetization in the leads \cite{ono97,bar98,bra99,kor99,tak98}. The
magnetoresistance ratio is defined by Eq.\ (\ref{tmr}). In order to
calculate the resistance of the FSET for arbitrary tunneling resistances the
tunneling Hamiltonian must be treated in a nonperturbative manner. This can
be realized via the path integral approach \cite{schon}. We use the standard
single-electron tunneling Hamiltonian with spin-dependent density of states
and tunneling matrix elements \cite{bar98,tak98} and limit our discussions
to the linear response regime.

At sufficiently high temperatures ($k_{B}T\gg E_{c}$) the system is in
the classical regime, the Coulomb charging effects are negligible and
the conductance of the system can be readily found. The dimensionless
conductance of the FSET for the parallel magnetic alignments is
$\alpha _{\text{P}}^{\text{cl}}=\alpha _{\text{M}}\left[
1+(1-P)(1-P^{\prime })/(1+P)(1+P^{\prime })\right] $, where $\alpha
_{\text{M}}$ is the dimensionless majority-band conductance of the
FSET with parallel magnetic alignments, and the spin polarizations of
the leads and island are denoted by $P$ and $P^{\prime }$,
respectively. Similarly, the dimensionless conductance of the FSET for
the anti-parallel magnetic alignments is $\alpha
_{\text{AP}}^{\text{cl}}=\alpha _{M}\left[ (1-P)/(1+P)+(1-P^{\prime
})/(1+P^{\prime })\right] $. The classical (high-temperature)
magnetoresistance ratio is
\begin{equation}
\gamma ^{\text{cl}}\equiv
\frac{R_{\text{AP}}^{\text{cl}}-R_{\text{P}}^{\text{cl}}}{R_{\text{P}}^{\text{c
l}}}=\frac{2PP^{\prime
}}{1-PP^{\prime }}\,.
\label{trc}
\end{equation}

At lower temperatures the Coulomb charging effects must be taken into
account. In this case, the resistance of the FSET can be calculated via
Kubo's formula \cite{wang}, 
\begin{equation}
\frac{R_{\eta }^{\text{cl}}}{R_{\eta }}=4\pi \lim_{\omega \rightarrow
0}\omega ^{-1}{\Im }m\{\lim_{i\omega _{l}\rightarrow \omega +i\delta
}\int_{0}^{\beta }d\tau e^{i\omega _{l}\tau }r(\tau )\}\,, 
\label{kubo}
\end{equation}
where the Matsubara frequencies are $\omega _{l}=2\pi l/\beta $ ($l$ is an
integral number) and 
\begin{eqnarray}
&&r(\tau )=Z_{\eta }^{-1}\sum_{k=-\infty }^{\infty }e^{2\pi ikn_{{\rm
ex}}} \nonumber \\ &&\times \int_{b_{k}}D\varphi
e^{-S_{\eta }[\varphi ]}\chi (\tau )\cos [\varphi (\tau )-\varphi
(0)]\,.  \label{rt}
\end{eqnarray}
where the damping kernel $\chi (\tau )$ is an even function with a
period $\beta $ and Fourier components $\chi (\omega _{l})=-|\omega
_{l}|/4\pi $ and $n_{\text{ex}}=C_{g}V_{g}/e$ is controlled by the
gate voltage. In Eq.~(\ref {rt}) the boundary condition for the
path-integral is $b_{k}\rightarrow \varphi (\beta )=\varphi (0)+2\pi
k$ and $Z_{\eta }$ is the partition function of the unbiased device,
namely a single-electron box (SEB)\cite{but}. $Z_{\eta }$ is readily
represented as a path integral \cite{schon,wang}
\begin{equation}  \label{z}
Z_{\eta }=\sum_{k=-\infty }^{\infty }e^{2\pi ikn_{{\rm
ex}}}\int_{b_{k}}D\varphi e^{-S_{\eta }[\varphi ]}\,.
\label{functional}
\end{equation}
The action is given by 
\begin{eqnarray}
&&S_{\eta }[\varphi ]=\int_{0}^{\beta }d\tau
\frac{\dot{\varphi}^{2}(\tau )}{4E_{c}} \nonumber \\
&&-\alpha _{\eta }^{\text{cl}}\int_{0}^{\beta }d\tau \int_{0}^{\beta
}d\tau ^{\prime }\chi (\tau -\tau ^{\prime })\cos [\varphi (\tau
)-\varphi (\tau ^{\prime })]\,, \label{action}
\end{eqnarray}
which depends on the relative orientations of the magnetizations in
the central grain and the leads through the second term in Eq.\
(\ref{action}) describing the contribution of the tunnel junctions. If
this term is neglected, the charging energy determines the action, and
the sequential tunneling processes are recovered. If one expands the
partition function (\ref{functional}) in terms of the dimensionless
conductance, a Taylor series is obtained containing sequential
tunneling, co-tunneling and higher order contributions. However in
practice, it is quite tedious to evaluate the path integral term by
term in this way. Instead, we will evaluate the DC current
nonperturbatively.

Note that the current-current correlation functions of the SET contain both
sine and cosine terms in the phase differences, as described in Ref.\ \cite
{wang}. However, using current conservation through the first and the second
junction it can be seen that the contribution from the sine term to the
conductance is proportional to the contribution from the cosine term and we
obtain the simplified expression (\ref{rt}).

Let us first discuss the most interesting case when the Coulomb
charging energy is much larger than the thermal energy and there is a
strong influence of the Coulomb charging on the current. In this case
the conductance of the FSET can be varied by tuning the gate voltage
\cite {ave91,wang}. We use Monte Carlo simulations to compute $r(\tau
)$~(\ref{rt}), and then calculate the conductances via the Kubo's
formula (\ref{kubo}).  Since in the Coulomb blockade regime, the TMR
for integer $n_{{\rm ex}}$ is at a maximum \cite{ono97,tak98}, we will
concentrate on this case to find the largest TMR and we set $n_{{\rm
ex}}=0$. The problem is reduced to the evaluation of expectation
values, which we will treat by Monte Carlo simulations \cite{weg}.

The computations have been carried out using the standard Metropolis
algorithm. The convergence of the simulations have been checked by
increasing the number of Trotter indices, which are the numbers of
slices in the imaginary time interval $[0,\beta ]$ and are found to
scale linear in $\beta $. Therefore at low temperatures where $\beta $
is large the simulations are time-consuming. Samplings have been done
every 5 passes, and the program has been ran for a sufficiently long
time to equilibrate the system before sampling. In order to guarantee
a high precision of the Monte Carlo simulations, the number of
samplings have been determined so that the result with doubled
sampling number $2N$ is the same as the result with sampling number
$N$. The typical number of samplings are found to be of order of
$10^{6}$ for the parameters described below. This procedure is
necessary because we calculate the Fourier components from the Monte
Carlo data and then carry out the analytical continuation (\ref{kubo})
using P\'{a}de approximate to get the conductance of the FSET
\cite{man90,cha95}.

When the tunnel resistances are close to the quantum resistance, the
co-tunneling result of the conductance of the SET
\cite{ave91,tak98,joy97} is recovered. In order to find out if higher
order tunneling processes will enhance the TMR, we show the result for
the conductance of a typical value in the strong tunneling case,
namely $\alpha _{M}=10$. We consider Ni leads with spin polarization
$P=0.23$, and Co island with $P^{\prime }=0.35$, as in the experiment
by Ono {\it et al} \cite{ono97}. For this device, the prediction of
the calculations of the TMR in the classical regime (high-temperature)
yields a value of 17.5 \% and in the Coulomb blockade regime the
co-tunneling formalism gives a value of 38 \% \cite{tak98} . The
computed results of the TMR as a function of $E_{c}/k_{B}T$ are shown
in Fig.~\ref{fig1}. In the low temperature quantum transport regime,
the TMR is denoted by $\gamma ^{\text{qu}}$. We see that the TMR is
strongly enhanced by lowering the temperature and reaches the
co-tunneling result \cite{tak98} when $E_{c}/k_{B}T\simeq 15$. In the
case of co-tunneling, the TMR saturates at this temperature and
remains constant at lower temperatures. We would like to emphasize
that in our case by including higher order tunneling processes a
further large enhancement of the TMR occurs at even lower
temperatures, e.g. when $E_{c}/k_{B}T\simeq 40$ the TMR is enhanced to
90 \%, which is a much larger value than in the co-tunneling
regime. At even lower temperatures a further enhancement is
expected. The fact that there is a large enhancement of the TMR in the
strong tunneling regime can be understood as a consequence of higher
order tunneling. In the co-tunneling regime the conductance of the
device is proportional to the square of the tunnel conductance which
roughly doubles the TMR with respect to the classical value. In our
case with $\alpha _{M}=10$ the conductance increases even faster than
the square of the tunnel conductance and hence the TMR is further
enhanced. The numerical results of the TMR in the low temperature
quantum regime can roughly be fitted by
\begin{equation}
\gamma ^{\text{qu}}=\gamma ^{\text{cl}}+\mu ^{\text{qu}}\left(
1+\gamma ^{\text{cl}}\right) (E_{c}/k_{B}T)^{2}\,, \label{trq}
\end{equation}
where we choose $\mu ^{\text{qu}}=0.0006$ so that the fitting curve is close
to most of the data. Note that although $\mu ^{\text{qu}}$ is quite small,
the second term on the right hand side of (\ref{trq}) can be as large as
unity when the temperatures are sufficiently low leading to a large
enhancement of the TMR. We have thus shown that there is a large enhancement
of the TMR in the strong tunneling regime at low temperatures, which is well
beyond the results from the sequential tunneling and co-tunneling formalisms.

Let us now consider the semiclassical regime when the thermal energy $k_{B}T$
is larger than the Coulomb charging energy $E_{c}$. The fluctuation
amplitude of the phase variable can then be estimated from the kinetic part
of the action giving a value of the order of $[E_{c}/k_{B}T]^{1/2}$, and the
dominating part of the action is Gaussian with some corrections from the
nonquadratic parts of the tunneling term in the total action. By expanding
the cosine function in the action Eq.~(\ref{action}) and in Eq.~(\ref{rt}), the
resistance of the FSET to the second order in $E_{c}/k_{B}T$ is 
\begin{equation}
\frac{R_{\eta }^{\text{se}}}{R_{\eta
}^{\text{cl}}}=1+\frac{E_{c}}{3k_{B}T}+(0.04-0.02\alpha _{\eta
}^{\text{cl}})\left( \frac{E_{c}}{k_{B}T}\right) ^{2}\,.
\label{rht}
\end{equation}
The calculation leading to the above equation is very similar to the
one of the nonmagnetic SET carried out in Ref.\ \cite{wang}. The two
first terms in Eq.~(\ref{rht}) have also been obtained in Ref.\
\cite{tak98}. The third term could also have been derived from the
formalism used in Ref.\ \cite {tak98}, but the last term is due to
higher-order tunneling. It is only this term that gives a
temperature-dependent contribution to the TMR at high
temperatures. The TMR in the semiclassical regime is
\begin{equation}
\gamma ^{\text{se}}=\gamma ^{\text{cl}}+\mu ^{\text{se}}\left(
1+\gamma ^{\text{cl}}\right) \left( E_{c}/k_{B}T\right) ^{2},\,
\label{trse}
\end{equation}
where  
\[
\mu ^{\text{se}}=0.04\alpha _{M}(1+PP^{\prime })/\left[ (1+P)(1+P^{\prime
})\,\right] .
\]
For $P=0.23$ and $P^{\prime }=0.35$, we find $\mu ^{\text{se}}\approx
0.013\alpha _{M}$ and the TMR for intermediate Coulomb charging energy
in the strong tunneling regime, e.g.\thinspace\ for $E_{c}=k_{B}T/4$
and $\alpha _{M}=10$, increases by less than 1 \% giving a TMR much
smaller than the low-temperature co-tunneling result of 38 \%. The
above analysis shows that when the thermal energy is dominating, the
Coulomb charging effect does not drastically change the TMR, and the
role of the higher order tunneling processes is not very important. In
order to make the TMR significantly different from the corresponding
values of a macroscopic device, one has to minimize the junction
capacitances so that the devices work in the low temperature regime.

It is interesting to note that the enhancement of the TMR the in low
temperature quantum regime~(\ref{trq}) has a similar
temperature-dependence as the enhancement of the TMR in the high
temperature semiclassical regime (\ref {trse}), with $\mu
^{\text{qu}}$ much smaller than $\mu ^{\text{se}}$. This can be
understood in terms of smearing of the Coulomb blockade by the strong
tunneling processes. Taking into account that at low temperatures the
renormalized charging energy of the FSET, $E_{c}^{\ast }$, is one
order of magnitude smaller than the bare charging energy $E_{c}$ when
$\alpha _{M}=10$ \cite{weg}, we see that $(E_{c}^{\ast }/k_{B}T)^{2}$
$\sim $ $0.01(E_{c}/k_{B}T)^{2}$, consistent with Eqs.~(\ref{trq}) and
~(\ref{trse}) where $\mu ^{\text{qu}}/\mu ^{\text{se}}\sim 0.01$. The
argument used here will fail at sufficiently low temperatures, where
the TMR can hardly be described by a power law of
$(E_{c}/k_{B}T)^{2}$. In such an extremely low temperature and strong
tunneling regime, more investigations need to be done.

In conclusion, we have derived formulas for the resistance of the
ferromagnetic single-electron transistor which are valid also in the regime
when the junction resistances are smaller than the quantum resistance. In
the strong tunneling regime we find that the magnetoresistance ratio is
largely enhanced by the higher order tunneling processes at low
temperatures. The magnetoresistance ratio in this regime is much larger than
the value predicted by only including sequential or co-tunneling processes.
We find that the TMR can at least be enhanced by a factor of 5 relative to
the classical value at low temperatures. When the thermal energy is larger
than the charging energy, the enhancement of the TMR is small, even when the
junction resistances are much smaller than the quantum resistance.

The authors are grateful to G. E.~W.~Bauer, K.-A. Chao, J. Inou,
A.~N.~Korotkov, Yu. V.~Nazarov and A. A.~Odintsov for stimulating
discussions and R.~Egger for valuable support of the Monte Carlo
simulations.  X.~H.~W.~would like to thank the Swedish TFR for
financial support and A.~B.~acknowleges financial support from the EU
TMR Research Network No.~FMRX-CT96-0089 (DG12-MIHT).


\begin{figure}[hbt]
\caption{The tunnel magnetoresistance ratio (TMR) as a function of the
dimensionless charging energy in the strong tunneling regime for
$\protect \alpha_M=10$. The diamonds are Monte Carlo data with errors
smaller than the sizes, and the solid line is calculated via Eq.~7 for
$\protect\mu^{\text{qu}}=0.0006$.}
\label{fig1}
\end{figure}

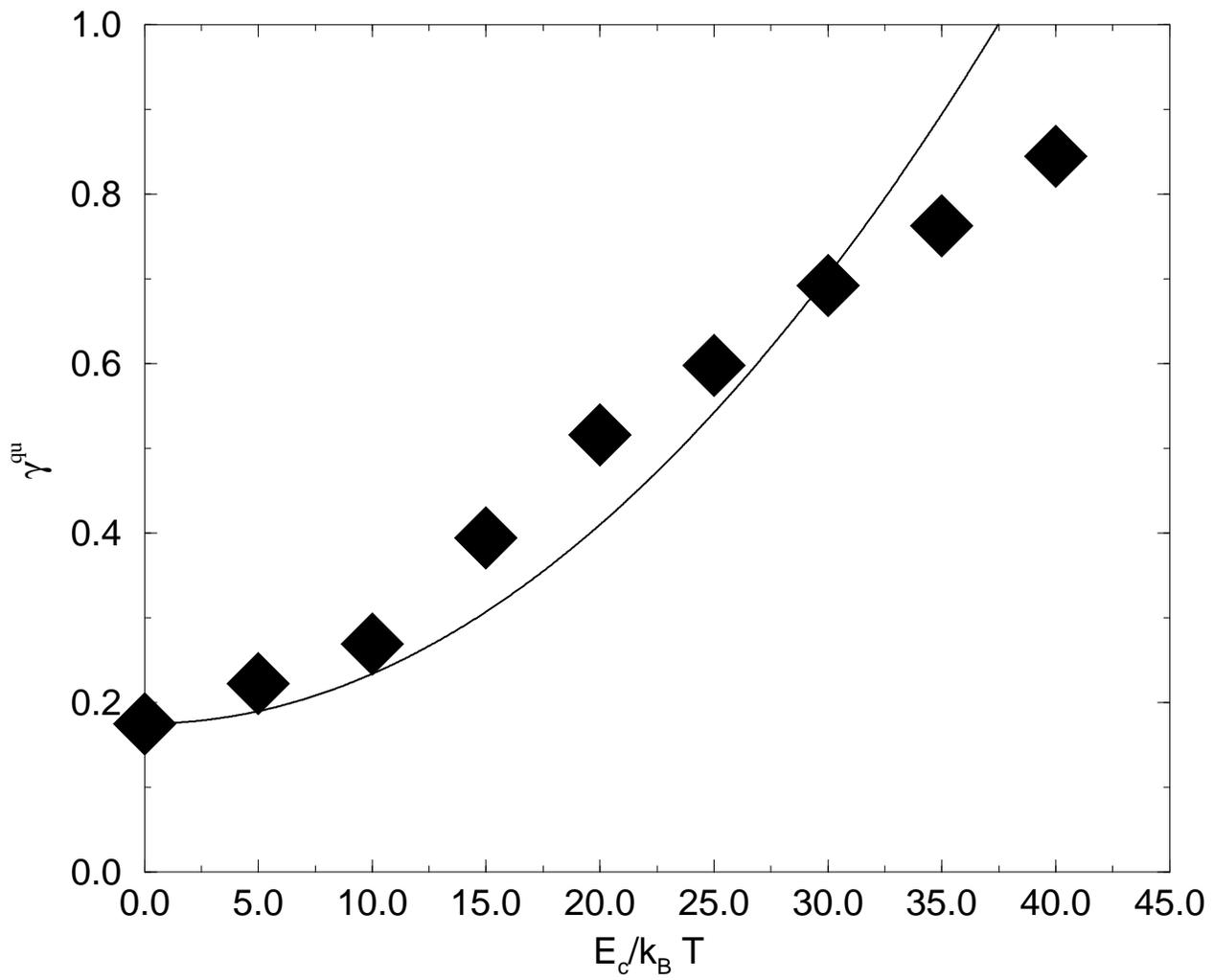
\end{document}